\documentstyle[aps,multicol]{revtex} 
\input{epsf}
\begin{document}
\draft
\title{Intrinsic resistivity and the SO(5)
theory of high-temperature superconductors}
\author{Daniel E.~Sheehy and Paul M.~Goldbart}
\address{Department of Physics and Materials Research Laboratory,\\ 
University of Illinois at Urbana-Champaign, Urbana, Illinois 61801, USA}
 \date{P-97-11-034-ii; December 5, 1997}
%
\maketitle
\begin{abstract}
The topological structure of the order parameter in Zhang's SO(5) theory
of superconductivity allows for an unusual type of dissipation mechanism
via which current-carrying states can decay.  The resistivity due to
this mechanism, which involves orientation rather than amplitude
order-parameter fluctuations, is calculated for the case of a thin
superconducting wire.  The approach is a suitably modified version of
that pioneered by Langer and Ambegaokar for conventional superconductors.
\end{abstract}
\begin{multicols}{2}
\narrowtext
Zhang's approach to the physics of the high-temp{\-}erature
superconductors~\cite{REF:Zhang} is rooted in the observation that the
phase diagram of these materials contains nearby regions of
antiferromagnetism and superconductivity.  This has led Zhang to 
propose a low-energy effective description of these materials that 
combines the U(1) symmetry sector associated with the
superconductivity with the SO(3) symmetry sector associated with  the
antiferromagnetism, SO(5) being the minimal symmetry group that admits
this combination.  In the resulting description, the state of the system
is characterized, locally, by a five-dimensional 
\lq\lq superspin\rlap,\rq\rq\ subject to a symmetry-reducing term that 
favors one or other of the sectors, U(1) or SO(3).  The superspin is
constrained to have unit magnitude, which is appropriate for
temperatures sufficiently low that fluctuations in the magnitude of the
superspin are negligible.  The possibility of rotations between the U(1)
and SO(3) sectors is where a number of the unusual consequences of Zhang's
approach lie~\cite{REF:arovas,REF:Demler,REF:hedgehog}.

It has long been appreciated~\cite{REF:Little,REF:LA} that in
conventional superconductors topologically accessible fluctuations in
the {\it amplitude\/} of the superconducting order parameter provide an
intrinsic mechanism via which supercurrent can be dissipated in a thin
wire.  A uniform current-carrying state characterized by a specific
uniform phase-gradient along the wire is only {\it metastable\/},
thermodynamically, so that by undergoing amplitude-reducing thermal
fluctuations the system can decrease its current.  The purpose of the
present Paper is to investigate a related intrinsic dissipation
mechanism appropriate for the superconducting state of the SO(5) model 
of high-temperature superconducting materials. 
In this case, the relevant dissipative process is a
fluctuation in the {\it orientation\/} of the superspin, during which
the system temporarily becomes antiferromagnetic along a small segment
of the wire, allowing the analogue of a phase-slip process to occur.
Such fluctuations are equivalent to the passing of superconducting
vortices with antiferromagnetic cores~\cite{REF:arovas} across the wire. 
Related dissipative processes have been addressed in the context of 
superfluid $^3$He-A~\cite{REF:Volovik} and (for appropriate values of 
the gradient coupling constants) thin tubes of nematic liquid
crystal~\cite{REF:Goldbart}. 
By constructing a version of the approach to intrinsic dissipation
pioneered by Langer and Ambegaokar for conventional
superconductors~\cite{REF:LA}, suitably modified for the SO(5) model of
superconductivity, we shall estimate the free-energy barrier for this
type of fluctuation and, hence, arrive at an estimate of the
current-voltage relationship for a sufficiently thin wire.

In Zhang's theory of high-temperature superconductivity and 
antiferromagnetism, then, the local state of the system at the 
spatial position ${\bf r}$ is determined by a five-component unit-vector 
field ${\bf n}({\bf r})$ such that the components
$n^1$, $n^2$ and $n^3$ together specify the antiferromagnetic
N\'eel vector, and the components $n^4$ and $n^5$ together specify the 
amplitude and phase of 
the superconducting order.  The constraint on the magnitude of ${\bf n}$
implements the notion that the enhancement of antiferromagnetic order is
necessarily accompanied by the diminution of 
superconducting order, and vice versa.
The free-energy density $f$ comprises an isotropic gradient term along
with a symmetry-reducing term:
\begin{equation}
f={\rho\over{2}}\sum_{a=1}^{5}\partial_{\mu}n^{a}\,\partial_{\mu}n^{a}
	-{g\over{2}}\sum_{a=1}^{3}(n^a)^2\,.
\end{equation}
Here, the (real-space) subscript $\mu$ runs from $1$ to $3$ 
(or $x$, $y$, $z$), 
repeated indices being summed over.  The (chemical-potential dependent)
parameter $g$ governs whether the stable homogeneous state is
superconducting ($g<0$), as we select here, or antiferromagnetic
($g>0$).  We parametrize ${\bf n}$ via the angles
$\{\tilde{\theta},\tilde{\phi},\tilde{\psi},\tilde{\chi}\}$ 
such that
\begin{eqnarray}
n^1=\sin\tilde{\theta}\cos\tilde{\psi}\cos\tilde{\chi},
&&\qquad
n^4=\cos\tilde{\theta}\cos\tilde{\phi},
\nonumber\\
n^2=\sin\tilde{\theta}\cos\tilde{\psi}\sin\tilde{\chi},
&&\qquad
n^5=\cos\tilde{\theta}\sin\tilde{\phi}.
\nonumber\\ 
n^3=\sin\tilde{\theta}\sin\tilde{\psi},\phantom{\sin\tilde{\chi}}\,\,
&&\qquad
\nonumber
\end{eqnarray}%
The angle $\tilde{\theta}$ measures the relative amount of
antiferromagnetic versus superconducting order (without regard for the
orientation of the antiferromagnetism or the phase of the
superconductivity), with $\tilde{\theta}\equiv 0$ in the purely
superconducting state.  Furthermore, $\tilde{\phi}$ is the phase of the
superconductivity, and $\tilde{\psi}$ and $\tilde{\chi}$ specify the
orientation of the antiferromagnetic N\'eel vector.

Let us suppose that the system consists of a wire of length $L$,
sufficiently long and narrow that we may assume that
$\{\tilde{\theta},\tilde{\phi},\tilde{\psi},\tilde{\chi}\}$ do 
not vary in the directions transverse to
the wire (i.e.\ in the $y$ or $z$ directions).  Under these
circumstances $f$ becomes
\begin{mathletters}
\begin{eqnarray}
f&=&(\rho/2)
 \big\{  (\partial_x\tilde{\theta})^2 
	+(\partial_x\tilde{\phi})^2\,\cos^2\!\tilde{\theta}
        +(\partial_x\tilde{\psi})^2\,\sin^2\!\tilde{\theta} 
\nonumber
\\&&\qquad
   	+(\partial_x\tilde{\chi})^2\,\sin^2\!\tilde{\theta}
         \cos^2\!\tilde{\psi}  \big\}
	+(\vert g\vert/2)\sin^2\!\tilde{\theta}
\label{EQ:lagrangian}
\\
&=&
(|g|/2)\left\{
	 (\partial_{\tau}\theta)^2
	+(\partial_{\tau}\phi)^2\,\cos^2\!\theta
 	+(\partial_{\tau}\psi)^2\,\sin^2\!\theta\right.
\nonumber
\\&&\qquad
  \left.+(\partial_{\tau}\chi)^2\,\sin^2\!\theta\cos^2\!\psi
        +\sin^2\!\theta\right\}.
\label{EQ:lagrangian2}
\end{eqnarray}%
\end{mathletters}%
To obtain Eq.~(\ref{EQ:lagrangian2}) we have exchanged the independent
variable $x$ for its dimensionless counterpart $\tau$, defined via
$\tau\equiv x/\xi_{\pi}$, where $\xi_{\pi}\equiv\sqrt{\rho/|g|}$ is the
correlation length for antiferromagnetic fluctuations, which sets the
length-scale for dissipative events.  Furthermore, we have defined the
function $\theta$ such that $\theta(\tau)\equiv\tilde{\theta}(x)$, and
similarly for $\phi$, $\psi$ and $\chi$. The quantity $\ell$ will denote
the dimensionless length of the wire, i.e.\ $\ell\equiv L/\xi_{\pi}$.

In order to estimate the rate at which current-dis{\-}sipating processes
occur we compute the height of the free-energy barrier opposing them.
To do this we follow Langer and Ambegaokar~\cite{REF:LA} and seek the
metastable (i.e.\ uniform, current-carrying) states between which the
system fluctuates, and the transition (i.e.\ unstable saddle-point)
states through which the system passes as current is dissipated.   Both 
classes of states, metastable and transition, are stationary 
configurations of the free energy, and therefore obey the corresponding
Euler-Lagrange equations.  
These equations may be simplified, however, due to the homogeneity 
(i.e.\ $\tau$-independence) and gauge-invariance in the superconducting 
sector (i.e.\ $\phi$-independence) of $f$.  The former symmetry leads 
to the existence of the first integral
\begin{mathletters}
\begin{equation}
\dot{\theta}^2
	\!+\!\dot{\phi}^2\,\!\cos^2\!\theta
        \!+\!\dot{\psi}^2\,\!\sin^2\!\theta
        \!+\!\dot{\chi}^2\,\!\sin^2\!\theta\cos^2\!\psi
	\!-\!\sin^2\!\theta\equiv\epsilon,
\label{EQ:FI}
\end{equation}
where overdots denote derivatives with respect to $\tau$, and $\epsilon$
is constant.  The latter symmetry leads to the existence of a
cyclic coordinate, so that
\begin{equation}
\dot{\phi}\cos^2\!\theta\equiv I\,.
\label{EQ:CyCo}
\end{equation}%
\end{mathletters}%
Here, the dimensionless supercurrent density $I$, in terms of which the
dimensionful supercurrent density $J$ is given by 
$J=2e\rho I/\hbar\xi_{\pi}$, is constant.  By replacing $\dot{\phi}$ 
in Eq.~(\ref{EQ:FI}) with $I$, using Eq.~(\ref{EQ:CyCo}), we obtain 
\begin{equation}
\dot{\theta}^2
	+{I^{2}\over{\cos^2\!\theta}}
        +\dot{\psi}^2\,\!\sin^2\!\theta
        +\dot{\chi}^2\,\!\sin^2\!\theta\cos^2\!\psi
	-\sin^2\!\theta=\epsilon\,.
\label{EQ:odeNOT}
\end{equation}

The relevant supercurrent-carrying metastable states are 
solutions of Eqs.~(3)
for which $\theta(\tau)\equiv 0$ (and therefore $\psi$ and $\chi$ play 
no role) and $\dot{\phi}=I$. We refer to these states as \lq\lq uniformly
winding\rq\rq\ because in them $\phi$ increases linearly with position
along the sample.  As for the transition states, we assume that they
possess appreciable antiferromagnetic order only within a small segment 
of the wire, and that elsewhere $\theta$ is 
negligibly small.  Furthermore, the orientation of the antiferromagnetic 
order generated during the fluctuation is uniform, because transition states 
with inhomogeneous antiferromagnetic orientation would have higher free 
energy, and would correspondingly occur less frequently.  
Therefore we only consider transition states for which 
$\dot{\psi}\equiv\dot{\chi}\equiv 0$.  Thus, by applying 
Eq.~(\ref{EQ:odeNOT}) far from the antiferromagnetic region we 
determine that 
\begin{equation}
I^2=\epsilon\,.
\label{EQ:eps1}
\end{equation}
Let us denote the extreme value of $\theta(\tau)$ by $\theta_0$, and let
us suppose that it occurs at the location $\tau=\tau_{0}$. At this
location $\dot{\theta}=0$, so that Eq.~(\ref{EQ:odeNOT}), combined with
Eq.~(\ref{EQ:eps1}), gives 
\begin{mathletters}
\begin{eqnarray}
\dot{\theta}^2&=&
\sin^2\!\theta-I_{\rm t}^2\tan^2\!\theta\,,
\label{EQ:dthetadt}
\\
I_{\rm t}^2&=&
\cos^2\!\theta_0\,. 
\label{EQ:currangle}
\end{eqnarray}%
\end{mathletters}%
By integrating Eq.~(\ref{EQ:dthetadt}) we find 
\begin{equation}
\sin\theta(\tau)=
\sin\theta_0\, 
\sqrt{1-\tanh^2[(\tau-\tau_{0})\sin\theta_0]}\,. 
\label{EQ:thetasol}
\end{equation}
Finally, by using Eq.~(\ref{EQ:CyCo}), along with 
Eq.~(\ref{EQ:thetasol}), we obtain 
\begin{eqnarray}
&&\phi(\tau)-\phi(\tau_{0})=
I_{\rm t}\,
\int_{\tau_{0}}^{\tau}d\tau'\big/\cos^{2}\!\theta(\tau')
\label{EQ:deltaphi}
\\
&&\quad=
I_{\rm t}(\tau-\tau_{0})
+\arctan\left\{\tan\theta_0
 \tanh\left[(\tau-\tau_{0})\sin\theta_0\right]\right\}. 
\nonumber
\end{eqnarray}%
\hbox{}
 \begin{figure}[hbt]
 \epsfxsize=\columnwidth
 \vskip-2.75truecm
 \centerline{\epsfbox{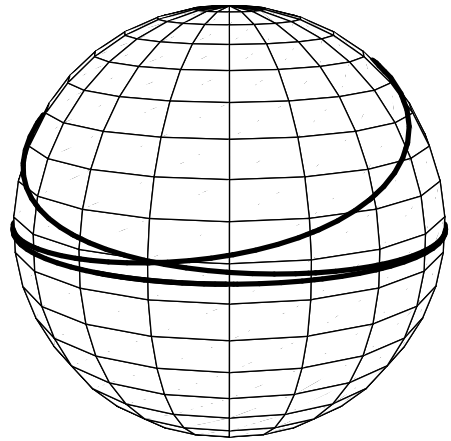}}
 \vskip-3.40truecm
\caption{Parametric plot of a typical (partially antiferromagnetic)
transition state (full curve) on the surface of the unit 2-sphere in the
space spanned by $\{n^4,n^5,\sin\theta\}$. Trajectories of the (purely
superconducting) metastable states lie on the equator of this sphere,
for which $\theta=0$.}
\label{FIG:one}
\end{figure}%
Thus we see that far from $\tau_{0}$ (i.e.\ far from the center of the
fluctuation) the transition states wind uniformly, as do the metastable
states.  However, in the transition states the order parameter undergoes
an orientational distortion over a length of order $\xi_{\pi}$ that
takes $\theta$ out of the $\theta=0$ plane.  Thus, a given transition 
state possesses a region of antiferromagnetic order in a fixed  
but arbitrary direction.  Indeed, there is
a family of symmetry-related transition states generated by changing
this direction, just as there is a family of transition states
generated by the (arbitrary) location of $\tau_{0}$ and the (arbitrary)
overall phase.  An example of a transition state is shown in
Fig.~\ref{FIG:one}. Hence we see that the relevant fluctuation for SO(5)
superconductors is one in which supercurrent can be dissipated via a
thermally activated process in which a \lq\lq loop\rq\rq\ of order
parameter passes over the order-parameter sphere, so that the
total phase difference (and hence supercurrent in the resulting
metastable state) is reduced by $2\pi$. Intuitively, it seems reasonable
that if the \lq\lq anisotropy\rq\rq\ $g$ is not too large (it being
adjustable by changing the chemical potential) then this type of
fluctuation should be less energetically costly than amplitude
fluctuations, and should hence provide the dominant pathway for
dissipation.  
Having determined the form of the metastable and transition states, we
now identify the particular transition state through which the system
passes as current is dissipated from a given metastable state. 
In passing from a metastable state to the corresponding transition
state, en route to diminishing the current, the superspin nowhere
becomes entirely antiferromagnetic (i.e.\ $|\theta|<\pi/2$). 
[The transition state represents a configuration in which a loop is 
about to pass over the order parameter sphere, but has not
yet done so.]\thinspace\ Thus, if we consider a transition from one
metastable state (which we refer to as the ``upper metastable state'')
to a metastable state with $2\pi$ less total phase difference (the
``lower metastable state'') then the actual loss of the phase will occur
in passing from the transition state to the lower metastable state.
Therefore, the total phase-difference across the sample $\Delta\phi$ is
the same in the transition state and the upper metastable state, so that
\begin{equation}
\int_0^{\ell}d\tau\,
\dot{\phi}_{\rm m}
  =  \Delta\phi
  =\int_0^{\ell}d\tau\,
\dot{\phi}_{\rm t}\,,
\label{EQ:implicit}
\end{equation}	
where $\phi_{\rm m}(\tau)$ is the phase of the uniform metastable state
and $\phi_{\rm t}(\tau)$ is the phase of the transition state. 
This formula provides a connection between the current in the upper
metastable state $I_{\rm m}^+$ and that in the transition state $I_{\rm t}$. 
The left-most term in this equation is given by $I_{\rm m}^+\ell$, which 
follows readily from the uniformly twisting character of the metastable 
states. The right-most term may be evaluated via Eq.~(\ref{EQ:deltaphi}), 
which gives $I_{\rm t}\ell+2\theta_{0}$, or equivalently, using 
Eq.~(\ref{EQ:currangle}), $I_{\rm t}\ell+2\arccos I_{\rm t}$. 
Thus we arrive at an implicit equation for 
$I_{\rm t}$ in terms of $I_{\rm m}^+$: 
\begin{equation}
\label{EQ:delta}
I_{\rm m}^+\,\ell- I_{\rm t}\,\ell=2\arccos I_{\rm t}\geq 0\,.
\end{equation}

Next we use Arrhenius rate-law considerations to develop the transition
rate.  To do this, we need expressions for the free energies of the
upper ($+$) and lower ($-$) metastable states, as well as of the
transition state connecting them.  From Eq.~(\ref{EQ:lagrangian2}) and
the form of the upper and lower metastable states we integrate over the
volume of the wire to obtain expressions for the free energies
$F^{\pm}_{\rm m}$ of the these states:
\begin{equation}
\label{EQ:uniff}
F^{\pm}_{\rm m}=(|g|/2)A\,\xi_{\pi}\,\ell\,\big(I^{\pm}_{\rm m}\big)^2\,.
\end{equation}
Here, $I_{\rm m}^{-}$ ($=I_{\rm m}^{+}-2\pi/\ell$) is the current in the 
lower metastable state and $A$ is the cross-sectional area of the wire.  
Similarly, the free-energy density of the transition states $f_{\rm t}$ 
may be obtained using Eqs.~(\ref{EQ:lagrangian2}), (\ref{EQ:CyCo}) and  
(\ref{EQ:dthetadt}):
\begin{equation}
\label{EQ:freee}
f_{\rm t}=(|g|/2)\,(I^2+2\sin^2\!\theta)\,.
\end{equation}
By integrating Eq.~(\ref{EQ:freee}) over the volume of the wire 
we arrive at an expression for the free energy $F_{\rm t}$ of the 
transition state:
\begin{mathletters}  
\begin{eqnarray}
F_{\rm t}&=&
(|g|/2)A\,\xi_{\pi}
\Big\{\ell\,I_{\rm t}^2
+2\int_0^{\ell}d\tau\,\sin^2\!\theta(\tau)\Big\}
\\&=&
\label{EQ:F_t}
(|g|/2)A\,\xi_{\pi}\,\Big\{\ell\,I_{\rm t}^2
+4\sqrt{1-I_{\rm t}^2 }\Big\}\,.
\end{eqnarray}%
\end{mathletters}%
By using Eqs.~(\ref{EQ:uniff}) and (\ref{EQ:F_t}) we obtain 
an expression for the free energy barrier $\Delta F$ for 
current dissipation,
\[
\Delta F\!\equiv\!F_{\rm t}-F_{\rm m}^{+}
={|g|\over{2}}A\xi_{\pi}\,
\!\Big\{\ell\,I_{\rm t}^{2}-\ell\,(I_{\rm m}^+)^{2}
\!+4\sqrt{1\!-\!I_{\rm t}^{2}}\Big\}\,, 
\]
which may be simplified by using the relation~(\ref{EQ:delta}) between 
$I_{\rm t}$ and $I_{\rm m}^{+}$.  By restricting our attention to 
states in which the phase winds many times along the wire
(i.e.\ $2\arccos I_{\rm t}\ll I_{\rm t}\,\ell$) we obtain
$I_{\rm t}^{2}-(I_{\rm m}^+)^{2}
	\simeq -4I_{\rm t}\arccos I_{\rm t}$, 
and thus $\Delta F$ becomes 
\begin{equation}
\label{EQ:fdiff}
\Delta F
=|g|\,A\,\xi_{\pi}\,\Big\{-2\,I_{\rm t}\arccos I_{\rm t}
+2\sqrt{1-I_{\rm t}^2}\Big\}\,.
\end{equation}
It should be noted that $I_{\rm t}=1$ is the critical current, in the 
sense that uniformly twisted states with larger values of the current 
are unstable rather than metastable.  
For later use, we note that for currents slightly smaller 
than critical $\Delta F$ in the SO(5) model is given by 
\begin{equation}
\label{EQ:df}
\Delta F\simeq 2\sqrt{2}\,|g|\,A\,\xi_{\pi}\,(1-I_{\rm t})^{3/2}\, ,
\end{equation}
whereas the corresponding expression for conventional 
superconductors~\cite{REF:LA} is 
\begin{equation}
\label{EQ:df1}
\Delta F\simeq \sqrt{2}\,(8/3)^{5/4}A\,\xi\,
(g_{\rm n}-g_{\rm s})\,
(1-I)^{5/4}\,.
\end{equation}
Here, ($g_{\rm n} - g_{\rm s}$) represents the free-energy cost of 
amplitude fluctuations, and is the analogue of the parameter $g$ 
within the present theory, and $\xi$ is the superconducting fluctuation 
correlation length.  

To compute the rates of current-decreasing and current-increasing
fluctuations we follow LA by assuming that these rates depend
exponentially on the free-energy barrier heights.  Specifically, the
rate $\Gamma(I_{\rm m}^+ \rightarrow I_{\rm m}^-)$ at which 
current-decreasing fluctuations occur is given by
\begin{mathletters}
\begin{eqnarray}
&&\Gamma(I_{\rm m}^+ \rightarrow I_{\rm m}^-)=
 \Gamma_0 \exp\{-\beta\,\Delta F)    
\\&& \quad
= \Gamma_0 \exp\left\{-\beta\,|g|\,A\,\xi_{\pi}
\left(-2I_{\rm t}\arccos I_{\rm t}
+2\sqrt{1-I_{\rm t}^2}\right)\right\} , 
\nonumber
\end{eqnarray}%
where $\beta=1/k_{\rm B}T$
measures the inverse temperature  and 
$\Gamma_0$ is an attempt frequency for dissipative 
fluctuations. Current-increasing fluctuations, in which the 
system passes from the lower metastable state to the upper
metastable state, also occur.  However, these fluctuations
have a higher barrier opposing them  (i.e.\ for them
$\Delta F\rightarrow \Delta F + F_{\rm m}^+ - F_{\rm m}^-$), and thus
occur at a lower rate.  By using Eq.~(\ref{EQ:uniff}) we therefore
obtain 
\begin{eqnarray}
&&\Gamma(I_{\rm m}^+ \leftarrow I_{\rm m}^-)= 
\\&& 
\Gamma_0 \exp\!\left\{\!-\beta|g|A\,\xi_{\pi}\!
\left(\!-2I_{\rm t}\arccos I_{\rm t}
+\!2\sqrt{1\!-I_{\rm t}^2}\!+2\pi I_{\rm m}^{+} \right)\!\right\}\!.
\nonumber
\end{eqnarray}%
\end{mathletters}%
The application of a voltage difference $\Delta V$ between the ends of the 
sample causes $\Delta\phi$ to increase linearly with time [i.e.\ 
$d(\Delta \phi)/dt=4\pi e\,\Delta V/h$, where $h$ is 
Planck's constant and $-e$ is the electronic charge]~\cite{REF:Jos}.  
In steady state, this will be balanced by the net rate $\Gamma_{\rm net}$ at 
which current-decreasing fluctuations occur, i.e.,    
\begin{equation}
\Gamma_{\rm net}=
\Gamma(I_{\rm m}^+\rightarrow I_{\rm m}^-)-  
\Gamma(I_{\rm m}^+ \leftarrow I_{\rm m}^-)= 
\frac{1}{2\pi}\frac{d(\Delta\phi)}{dt}\,.
\end{equation}
The difference between $I_{\rm t}$ and $I_{\rm m}^+$ is of order
$\ell^{-1}$, and we shall henceforth neglect it; thus we simply
refer to the current $I$.  This leads to the following expression for
the voltage between the ends of  the wire in terms of the 
current along it:
\begin{eqnarray}
&&\Delta V=
(h/e)\,\Gamma_0\,\sinh\{\pi\beta|g|A\xi_{\pi}\,I\}
\label{EQ:delv}
\\
&&\quad\times
\exp\left\{-\beta|g|A\,\xi_{\pi} 
\left(-2I\arccos I+2\sqrt{1-I^2}+\pi I\right)\right\}\,.
\nonumber
\end{eqnarray}%
The corresponding expression for conventional superconductors 
may be found in Eq.~(2.12) of Ref.~\cite{REF:LA}.
As the length-scale of the fluctuation is $\xi_{\pi}$ it is natural for 
$\Delta V $
to depend exponentially on $\beta |g|A\xi_{\pi}$ (i.e.\ the condensation 
energy in a correlation length per $k_{\rm B}T$).  Following McCumber 
and Halperin~\cite{REF:MH}, we estimate $\Gamma_0$ to be of order $N/\tau$, 
where $N\equiv 4\pi L/\xi_{\pi}$ reflects the range of possible locations 
along the wire at which the fluctuation may occur ($L/\xi_{\pi}$), 
as well as the variety of possible N\'eel-vector
orientations ($4\pi$).  For conventional superconductors $\tau$ diverges as
$T\rightarrow T_{\rm c}$.  For the SO(5) model we expect $\tau$ to
diverge as the chemical potential approaches its critical value.

It may be  useful to compare expression~(\ref{EQ:delv}) for $\Delta V$ 
with that obtained in Refs.~\cite{REF:LA,REF:MH} for the case of 
conventional superconductors.   
Within the SO(5) model the expression for $\Delta V$ in terms of 
the dimensionful supercurrent density $J$ is given, for 
$J\simeq J_{\rm c}$, by  
\begin{eqnarray}
&&\Delta V\simeq
(h/e)\,\Gamma_0\,
\sinh\{\beta AhJ/4e\}  
\label{EQ:VforSOF}
\\
&&\times
\exp\left\{-\beta A\left((hJ/4e) + 
 2\sqrt{2}\,|g|\,\xi_{\pi}\,(1-J/J_{\rm c})^{3\over{2}}\,\right)\right\}\, ,
\nonumber 
\end{eqnarray}%
whereas the corresponding expression obtained in 
Refs.~\cite{REF:LA,REF:MH} is given, for $J\simeq J_{\rm c}$, by 
\begin{eqnarray}
&&\Delta V \simeq
(h/e)\,\Omega_0 \,
\sinh\{\beta AhJ/4e\}  
\label{EQ:VforLA}
\\
&&\times\!
\exp\!\left\{\!-\beta\!A\!\left(\!(hJ/4e) \!+ 
 \!\sqrt{2}(8/3)^{5\over{4}}\xi
(g_{\rm n}\! -\! g_{\rm s})
(1\! -\! J/J_{\rm c})^{5\over{4}}\right)\!\right\} .
\nonumber 
\end{eqnarray}%
(We have ignored the current dependence of the attempt frequency 
$\Omega_0$ obtained in Ref.~\cite{REF:MH}.)\thinspace\ 
We see that in this regime the primary qualitative distinction between 
formulas~(\ref{EQ:VforSOF}) and (\ref{EQ:VforLA}) comes from the 
distinction between the nature of the fluctuations, and the resulting 
difference between the barrier heights. 

We conclude by noting that for conventional superconductors 
LA~\cite{REF:LA} have obtained, as a condition for the validity of 
their approach, the constraint that the wire be thinner than the 
temperature-dependent superconducting correlation length $\xi$.
(To accomplish this they consider the second variation of the 
free energy, and require that the transition state be unstable 
in only one direction in configuration space.)\thinspace\  We 
emphasize that the corresponding condition in the context of 
the SO(5) theory of high-temperature superconductivity is that 
the wire be thinner than the correlation length for orientation 
fluctuations of the superspin, i.e., $\xi_{\pi}$.  This length 
$\xi_{\pi}$ is expected to diverge when the chemical potential 
approaches the superconductor-to-antiferromagnet phase boundary.  

\smallskip
\noindent
{\it Acknowledgments\/} ---  
Helpful discussions with Erich Mueller and Yuli Lyanda-Geller are
gratefully acknowledged. This work was supported by the U.S.~Department
of Energy, Division of Materials Sciences, under Award
No.~DEFG02-96ER45439 through the University of Illinois Materials
Research Laboratory (DES), and by the U.S.~National Science Foundation
through grant DMR94-24511 (PMG).

\end{multicols}
\end{document}